\documentclass[]{aastex631}

\begin{document}

\title{Forbidden lines are not quenched.}

\author[0000-0001-6815-4055]{Jason Spyromilio}
\affiliation{European Southern Observatory \\
Karl-Schwarzschild-Str 2 \\ Garching D-85748, Germany}





\begin{abstract}

There is no such thing as ``quenching'' of forbidden lines.  

\end{abstract}

\keywords{History, Atomic physics}

\section{Introduction, Historical note \& Background} \label{sec:history}

In \cite{2006agna.book.....O} it is stated that: {\em when collisional de-excitation begins to be important, the cooling rate at a given temperature is decreased ...} and further  that {\em At high electron densities, collisional de-excitation can appreciably modify the radiative cooling rate ... }. The question arises:  How does this modification take place? In the same tome, \cite{2006agna.book.....O} it is noted that: {\em as [the free electron density] $N_e \rightarrow \infty$, [the cooling rate] $L_C \rightarrow N_l (g_u/g_l) exp(-\chi/kT) A_{ul} h \nu_{ul}$ [recovering] the thermodynamic equilibrium rate.}

Thus, within a few pages of the same book the authors present different behaviours of forbidden lines. Surely the thermodynamic rate is the highest rate. Taking a thermodynamical view, one cannot exceed, modulo population inversions, the Boltzmann population of atoms in an energy level. Therefore, the emission by a forbidden line surely is maximum at high electron density (caveats about optical depth apply). 

\cite{1927Natur.120..473B} identified the lines attributed to Nebulium as emission by transitions from what he described as `metastable' energy levels. The genius of Bowen\footnote{Bowen was quite a remarkable scientist: In addition to determining the source of emission lines in nebular spectra, he also commissioned the 200-inch telescope, invented the Bowen image slicer for spectrographs, identified most of the forbidden lines in the visible spectrum, found the mechanism for fluorescence in Helium, was director of Palomar, and much else.} in identifying the lines as emission by `forbidden' transitions is without doubt one of the most impressive breakthroughs in astronomy. However, the explanation in \cite{1927Natur.120..473B} and explicitly in \cite{1927PASP...39..295B} that: {\em If the mean life of the excited state before spontaneous emission is very long, as in the case of a metastable state, and if the mean time between impacts is short ... practically all of the atoms will return by the second process and no radiation will take place} has been the cause of some confusion and was discussed at the time. Unfortunately, it also appears to have taken root in astronomy as the `correct' explanation. 

Almost immediately, following the Bowen publication, \cite{1928Natur.121..618M} commented on the density issue. In \cite{1938Nat..142..644} it is clearly stated  that: {\em Part of the [Bowen] misunderstanding may arise from the {\bf \em erroneous belief} that if the probability of collisional de-excitation is greater than the probability of emission the atoms do not have the time to radiate. This reasoning would imply a difference between an atom that arrived in a metastable state 10$^{-8}$ seconds ago and one that may be have existed in that level for some seconds. The argument would imply that the quantum equation \footnote{In this text, including in direct citations from the literature, for consistency, we shall use the nomenclature of referring to $u$ for the upper energy level and $l$ lower. Where, in the quoted literature 1s and 2s ($a$ and $b$, $m$ and $n$) were used to denote lower and upper energy levels, they have been replaced herein with $u$ and $l$ to avoid confusion. One exception is when multiple levels are discussed in which case 0, 1, 2, 3 ... are used as indices, where 0 is then the ground state. Irrespective of the terms used by the authors, the symbols for collision strengths are $C$, and for cooling rates $L$. Thankfully it appears that the term A-value has survived.}, $I=N_u A_{ul} h\nu$ is wrong. [Contrarily] The intensity depends solely on the population and atomic constants, and the highest intensity occurs when $N_u$ is highest} and in \cite{1969MSRSL..25..113M}:
{\em The number of transitions, per second, from a volume containing $N_u$ atoms in level $u$ is $N_u A_{ul}$, {\bf \em irrespective of how long the atoms have been in level $u$}. And $N_u$, for a given mass, will be greatest when collisions are more frequent} and {\it the commonly expressed view that an atom cannot radiate if the inelastic collision frequency exceeds the Einstein transition probability is {\bf \em completely wrong}}.

Unfortunately, textbooks, lecture notes, and papers, have been discussing the effect of increasing free electron density using language that is vague, and  a few examples can be found the Appendix A. 

\section{The two level atom}

In an ensemble of two level atoms with lower {\em l} and upper {\em u} energy levels separated by $E_{ul}$ (cm$^{-1}$), bound electrons can be in either of these two levels and these electrons can move between the energy levels by collisional or radiative processes. The rate at which a spontaneous emission process takes place is given by the Einstein $A_{ul}$ (s$^{-1}$) coefficient. The energy of the photons emitted is $h \nu_{ul}$.

Electrons in the lower energy level can be collisionally excited to the higher level by interacting with free electrons or other colliders. Concentrating on collisions only by free electrons, the rate at which these excitations occur is the product of the free electron density $N_e$ (cm$^{-3}$) and the collision coefficient $C_{lu}$ (s$^{-1}$cm$^3$). Multiple textbooks describe the atomic physics that underpin the dependence of $C$ on temperature (T$^{-0.5}$). For our purposes here it is only necessary to note that calculations of the populations depend on both density and temperature. 

An atom with an electron in the upper energy level can also be collisionally de-excited should the electron move from the upper to the lower energy level using the same collisional mechanism operating in the opposite direction and the coefficient is then $C_{ul}$. The rate coefficient up is connected to the rate coefficient down by:

\begin{equation}
    C_{lu} = \frac{g_u}{g_l}C_{ul}e^{-E_{ul}/kT}
    \label{eq:culclu}
\end{equation}

where $g$ is the statistical weight of the energy level. This equation is important and we shall return to it later. The exponential term comes from the fact that to excite there is a minimum energy requirement on the free electron whereas to de-excite is cheap. However, any energy expended by the free electron gas to excite a bound electron is returned to the free electron gas in the downwards collision. Collisions, at least as defined herein, do not cool the gas. The energy that is removed from the gas is in the conversion of kinetic energy of the free thermal electron gas into photons that can escape the gas.

The bound electrons can also be radiatively excited and de-excited by coupling to the radiation field and the rates for these processes are given by the Einstein $B_{ul}$ and $B_{lu}$ coefficients. We shall ignore these processes for the time being. 

All of these processes take place irrespective of the high or low value of any of these rates. Forbidden and allowed lines obey the same laws of radiation, collisions and statistics. 

The metastable energy levels can also be populated direct by recombinations or cascades from upper levels or `sideways' radiative or collisional processes, and in the case of high density even self-absorption. However, for most of this note we shall focus on collisional excitation by a free thermal electron gas.

\subsection{Balance}

As we are currently ignoring the $B$ terms, the total rates at which electrons move up and down in an ensemble of atoms are thus: 

\begin{itemize}
\item Lower to upper: $N_e N_l C_{lu}$ 
\item Upper to lower: $N_e N_u C_{ul} + N_u A_{ul}$ 
 \end{itemize}

where $N_e$ is the free electron density, $N_l$ and $N_u$ (cm$^{-3}$) are the populations of atoms with electrons in the lower and upper energy level respectively. While in principle there need not be any equilibrium, since down rate only occurs when $N_u$ is not zero and $N_u$ only becomes non-zero if the  lower-to-upper transitions occur it is safe to write 

\begin{equation}
    N_e N_l C_{lu} = N_e N_u C_{ul} + N_u A_{ul}
    \label{eq:bala}
\end{equation}

and thus the flux in the transition $u \rightarrow l$ is

\begin{equation}
    N_u A_{ul} h \nu_{ul} = N_e N_l C_{lu} h \nu_{ul} \left( \frac{1}{1+\frac{N_e C_{ul}}{A_{ul}}} \right) \, {\rm (erg/s/cm^{-3}})
    \label{eq:nahnu}
    \end{equation}

\section{Limiting cases}

\subsection{Low electron density}

In the low electron density regime $N_e C_{ul}$ will much smaller than $A_{ul}$ and in equation\,\ref{eq:nahnu} the $N_e C_{ul}/A_{ul}$ term is near 0, and the term in the parenthesis is 1. Thus, the commonly used approximation for the radiation in the low density regime as can be found in \cite{2006agna.book.....O} and elsewhere is:

\begin{equation}
    L_{Cooling} = N_e N_l C_{lu} h \nu_{ul}\, {\rm (erg/s/cm^{-3}})
        \label{eq:cool}
\end{equation}

This approximation is very convenient as, assuming we can estimate $N_e$ from a line ratio and measure the flux of a line, then we can directly derive the $N_l$ which, in the case of a transition down to the ground state, where in a low density environment most atoms are found, is effectively the density of the ion. Simple integration gives total ion abundance. It is brilliantly easy, effective, and largely correct. The only atomic data required is a collision strength that can be found in the literature. However, it remains an approximation and does not reflect the physics of what is going on. Furthermore, extrapolating from this has caused much confusion. Collisions up and down are taking place with their respective rates and the correct formula to use is equation\,\ref{eq:nahnu}. 

A common phrasing appearing in the literature and discussions postulates that {\em there are so few free electrons available to de-excite the electron in the upper level that an electron {\bf \em spends enough time} there that it has a chance to spontaneously de-excite.} This is very confusing language. Alternatively the wording is: {\em because radiative de-excitation is so rare, a population of atoms in $N_u$ can be established} which is accurate but still can lead one to the concept of an individual electron hanging around waiting for something to happen to it. 

\subsection{Critical density}\label{critden}

The concept of a critical density has also proven to be very useful. The critical density is deemed to be the density at which the approximation in equation\,\ref{eq:cool} no longer holds. When $N_e$ is high enough that the collisional rate down competes with the radiative rate down, this is called the critical density. Typical values for the critical density range from 10$^6$\,cm$^{-3}$ for optical forbidden lines with A values of order 10$^{-2}$\,s$^{-1}$, to 10$^{11}$\,cm$^{-3}$ for UV intercombination lines with A values of order 100\,s$^{-1}$, to 10$^{15}$\,cm$^{-3}$ for permitted lines with A-values of order 10$^8$\,s$^{-1}$. We can thus define the critical density to be $N_c=A_{ul}/C_{ul}$ and can rewrite equation\,\ref{eq:nahnu} as:

  \begin{equation}
    N_u A_{ul} h \nu_{ul} = N_e N_l C_{lu} h \nu_{ul} \frac{N_c}{N_e + N_c}
    \label{eq:nahnc}
    \end{equation}  

The confusion arises here. It is oft stated that around the critical density the forbidden lines are {\em quenched} or {\em the electron {\bf \em does not spend enough time} in the upper energy level for it to get a chance to spontaneously decay} or {\em above the critical density the forbidden lines are less efficient at cooling} or {\em the density of the gas is high  enough and there is a high chance of collisional de-excitation {\bf \em before} a radiative decay can take place.}

There is no magic density, nor A-value, at which the physics of the atom changes. There is a density at which various approximations used to accelerate the computation of cooling break down. As the density increases the emissivity of the line increases, until it saturates.

\subsection{Densities higher than critical}

Above the critical density the population of atoms in the collisionally excited energy level asymptotically reaches the Boltzmann distribution. Equation\,\ref{eq:nahnu} is of a simple $ax/(1+bx)$ form. The limit as x goes to infinity is $a/b$. In that limit equation\,\ref{eq:nahnu} reduces to: 

\begin{equation}
    N_u A_{ul} h \nu_{ul} = \frac{N_e N_l C_{lu} h \nu_{ul}}{\frac{N_e C_{ul}}{A_{ul}}}
\end{equation}

and eliminating terms of both sides you end up with

\begin{equation}
    N_u = N_l\frac{C_{lu}}{C_{ul}}
    \label{eq:nunl}
\end{equation}

which evidently, combined with equation\,\ref{eq:culclu}, gives 

\begin{equation}
    N_u = N_l \frac{g_u}{g_l}e^{-E_{ul}/kT}
    \label{eq:nunl2}
\end{equation}

recovering the Boltzmann population for atoms. Thankfully this is no surprise. In thermal equilibrium detailed balance holds and for each process the rates up and rates down are equal and this is reflected in most texts. E.g. \cite{1969MSRSL..25..251F} write 
{\em [as] $N_e \rightarrow \infty $: then spontaneous radiative decay is negligible compared with collisional de-excitation and a Boltzmann distribution is obtained}. 

\subsection{Non-trivialities and where confusion reigns.}

Equation\,\ref{eq:bala}, in the full thermodynamic equilibrium limit (equation\,\ref{eq:nunl}), would appear to require that $N_u A_{ul}$ is zero and this is irrespective of the value of $A_{ul}$. This would apply to H$\alpha$ as much as to [O{\small III}] $\lambda\lambda$\,5007. This is the cause of some confusion. However, we should note that the balance equation for full thermodynamic equilibrium is: 

\begin{equation}
    N_l (N_e C_{lu} + B_{lu} U(\nu_{ul})) = N_u (N_e C_{ul} + A_{ul} + B_{ul} U(\nu_{ul}))
\end{equation}

where $U(\nu)$ is the radiation field and in full thermodynamic equilibrium the detailed balance also requires:

\begin{equation}
    N_l B_{lu} U(\nu_{ul}) = N_u A_{ul} + N_u B_{ul} U(\nu_{ul})
\end{equation}

In thermodynamic equilibrium there is no heating nor cooling. All processes are balanced by their inverse. Spontaneous emission in the full thermodynamic equilibrium is balanced by the $B$ terms. Spontaneous emission is not quenched by collisions. 

\section{So what is going on in the very high density regime}

In the very high density regime, and assuming the free electron gas has itself a thermal distribution, the number of atoms with bound electrons in the various energy levels follows a Boltzmann distribution reflecting the temperature of the free electron gas. The atoms are in equilibrium with the thermal bath. No cooling or heating is taking place. 

Since a thermal `bath' is never in isolation, its surface does radiate. It likely radiates as a black body. A black body being a perfect emitter, presupposes that no lines can be emitted from within. The atoms are in equilibrium. All photons emitted inside the emitting surface are re-absorbed. They are not `quenched' and it matters not whether the line is permitted or allowed. The difficulty in the language and interpretation comes in conditions of partial thermodynamic equilibrium. Obviously one energy level can have reached a Boltzmann population while another not. What happens to the atoms that have electrons in those energy levels? It would appear to violate the laws of quantum mechanics to declare that they do not offer the opportunity to the electron to spontaneously decay.

\section{Why is this confusing?}

The confusion seems to arise from a awkward transition in how we talk about the physics. A single atom clearly obeys the quantum mechanical laws about the probabilities of individual events occurring. In all of the discussion above we are dealing with a single bound electron. A single bound electron is either in $N_l$ or $N_u$. That single electron has a given probability to do something, either: it can be collisionally de-excited or it may spontaneously decay. 

Two ways of talking on this issue are presented below as they seem to reflect the two interpretations of the issue in conversations with one's peers. They are, first the [Bowen] view shared by a significant fraction of the community as sampled by the author, and second the minority [Menzel] view:

\begin{itemize}

\item {\bf The electron decides its fate}

Assume an electron has made it to the upper level and the transition down has an A-value of 10$^{-8}$\,s$^{-1}$. The idea that the electron would have to wait for 10$^8$\,seconds to be sure to decay seems intuitively correct. Thus if something else happened during those 10$^8$\,seconds to bump the electron out of the energy level (e.g. a downwards collision) then no radiation would be observed. So the actual flux of the line would decrease as collisions increased. In fact in the limit of high density the electron would have no chance be spontaneously decay. The original \cite{1927PASP...39..295B} statement.

If one interprets de-excitation as a competition between collisions and radiation, then clearly as the electron density increases the collisions win. In the limiting case of very high density there is no radiation. The electron that found itself in the upper energy level will always get wacked down before it gets a chance to do anything else. At low electron densities, the only action is $N_u A_{ul}$ and thus only photons, at intermediate electron densities the action is split between $N_e N_u C_{ul}$ and $N_u A_{ul}$, and at high electron densities the only action is $N_e N_u C_{ul}$ and thus no photons. 

\item {\bf Statistics rules}

Alternatively, the A-value of $10^{-8}$\,s$^{-1}$ does not imply that the electron has to `wait' for 10$^8$\,seconds to get a chance to radiate. The A-value implies that if we look at the energy level at random at any one time within a period 10$^8$\,seconds, should there be an electron there, it will spontaneously  decay by radiating a photon. So what matters is not the relative probability of something happening but rather the probability that an electron happens to be in the upper level when we look. That probability is highest when the energy level is occupied according to the Boltzmann distribution. 

As the density increases not only do the down collisions increase but so do the upwards collisions. The rate in fact is such that in the high density limit the population of the upper level is the Boltzmann one. Therefore, any line ratios will be dictated by the values of the statistic weights, excitation temperatures, and A values for each energy level and transition. 

Moreover, the emissivity of the individual line will be maximum at high density (modulo optical depth and other effects).

\end{itemize}

\section{So what about the cooling of a gas?}

It is perfectly logical, in the absence of a radiation field, and in the low density regime, to calculate the cooling rate as the difference between the collision rate up and down. At the high density regime one can end up subtracting very large numbers which computationally can be inaccurate but that is not the issue at hand here. Many codes indeed calculate cooling independently of line emission. As you reach Boltzmann distributions the collisional rate up will tend to equal the collisional rate down appearing to reduce the cooling rate. 

This does {\bf not} imply that the lines are ``quenched'' or collisionally supressed due to an increase in downward collisions. 

A gas with long optically thin path lengths can cool very effectively via collisionally excited forbidden lines. In fact it will do this at its best when the density is above the critical density. That's when the forbidden lines will be at their strongest. Typically such cooling, is aided by the fact that bulk motions in the gas are large enough that emission from one region is out of resonance with another. Contrarily an optically thick, high density gas, cools by the emission of photons from its surface. This may be less effective not because this emission is less efficient, in fact it is maximally efficient if it reaches the black body level, merely that the only participating atoms are at the surface and as opposed to the entire volume. 

\subsection{How come one sees both forbidden lines and allowed lines from the same gas?}

The concept of departure coefficients is very useful here. Departure coefficients are the difference between the actual population of level from the thermodynamic equilibrium population: $b_n = N_n / N^*_n$ where $N^*_n$ is the population of level $n$ in the Boltzmann limit.  In the absence of a radiation field ($U(\nu)$ being negligible) we get the ratio of departure coefficients (equation 4-14 \cite{1998ppim.book.....S})

\begin{equation}
\frac{b_u}{b_l} = \frac{1}{1+\frac{A_{ul}}{N_e C_{ul}}}
\end{equation}

Obviously, as has been oft stated herein, as $N_e \rightarrow \infty$ the ratio tends to 1. I.e. no departure from a Boltzmann distribution. Any particularly interesting physics occurs earlier. The departure coefficient for an allowed line at nebular densities is tiny (or enormous depending on how one sees it). Thus, allowed lines appear faint compared to a Boltzmann distribution for the same electron gas temperature. The departure coefficients for energy levels radiatively depopulated solely by forbidden lines may be significantly closer to 1. So the emissivity of the forbidden and allowed lines can be comparable. 

\section{Absorption of photons by electrons in metastable energy levels}

In a brilliant paper \cite{1968ApJ...152..701B} noted that the fine structure components of the ground state energy level of C{\small I}, Si{\small I} and other species can be populated by collisions even in an interstellar medium with densities of a few 100\,cm$^{-3}$. This in turn implies that absorption lines arising from allowed transitions connecting upper energy levels to these fine structure components of the ground state would be seen and could be used to establish the conditions in the ISM. These absorption lines are observed, for example in \cite{2004A&A...415...77H}.

In a much less brilliant work, \cite{1991ESOC...37..423S, 1991MNRAS.253P..25S} showed that the evolution of the optical depth of the forbidden $\lambda\lambda 6300,6363$ [O{\small I}] transitions can, in the expanding ejecta of supernovae, explain the observed evolution of the line ratio of the two lines from 1:1 (optically thick regime) to 3:1 (optically thin regime). In the ejecta of the supernova the ground state is internally in its Boltzmann distribution (through collisions). The two oxygen lines arising from the same upper level but connecting to different energy levels within the ground state of neutral oxygen have different optical depths as they are a function of their respective A values and the density of the gas. 

The purpose of this section is simply to point out that atoms with electrons that have arrived in an energy level via collisions exhibit all the normal radiative behaviours, including absorption of photons. Had there been `no time' to do so because collisions were busily taking the electrons away, this behaviour could not be exhibited. The devil's advocate argument could be that: the processes described here are of course upwards absorptions so it could be that the electron can radiatively go up but not down. That would appear very odd. 

\section{Conclusion}

An energy level population is only meaningful in the context of an ensemble of atoms. Thus, when the ensemble of atoms is in the high density collision dominant regime there are, {\bf always and at all times}, atoms with electrons in the N$_1$, N$_2$, N$_3$, ... states and the numbers of atoms in those states follow the Boltzmann statistics relative to the atoms with electrons in ground state $N_0$. These populations are established by thermalising the atom via collisions with the free electron gas or via other collisions. 

As \cite{1935tas..book.....C}  state {\em The actual emitted intensity of the component from state $u$ to state $l$ in ergs/second is equal to  $I_{ul}=N_uA_{ul}h\nu_{ul}$ where $N_u$ is the {\bf \em the number of atoms in state $u$ ...}} and \cite{1916DPhyG..18..318E} in the original derivation of the coefficients states {\em $N_u$ die Anzahl der Moleküle vom Zustand $Z_u$ bedeutet} translated in the Einstein Papers as {\em $N_u$ {\bf \em is the number of molecules} in state $Z_u$}.

In the high density limit a Boltzmann distribution of energy level populations is established and there is no suppression of the spontaneous emission. Downward collisions do not quench (collisionally suppress) emission lines. There is only a saturation of the flux of photons arising from spontaneous emission at its maximum value $I(\nu_{ul}) = A_{ul} h \nu_{ul} N_l (g_u/g_l)exp(-\chi_{ul}/kT)$ where $l$ is now the ground state ($E_l = 0$\,cm$^{-1}$).
\begin{acknowledgments}
Discussions with Stephane Blondin, Roberto Gilmozzi, Claes Fransson, Shri Kulkarni, Eline Tolstoy, Filippo Fraternalli, Bruno Leibundgut, Adam Rubin, Phil Pinto, Nando Patat, Michael Hilker, and many of the staff \& visitors, graduate students, and postdocs at ESO are gratefully acknowledged. John Raymond, the referee, was instrumental in improving this article. This research has made use of the Astrophysics Data System, funded by NASA under Cooperative Agreement 80NSSC21M00561. No funding agency was burdened by this work. 

\end{acknowledgments}

\bibliography{PASPsample631}{}
\bibliographystyle{aasjournal}

\appendix

\section{Quotations from the literature}\label{sec:quote}

This is a very small subsample of words (not quite random and certainly not exhaustive) and by default in some cases unfair on the authors selected. It is almost true to say that not a single incorrect word is present in the texts. It is the ambiguity in the wording that is the root of the confusion. 

\begin{itemize}

\item \cite{2007ASSL..342.....K}: {\it \begin{enumerate} \item $N_e \ll N^c_e$ (rarefied gas). Since $A_{ul} \gg C_{ul}N_e$ we have [the photon rate] $Q_{ul} \approx N_l N_e C_{ul}$ ... All of the atoms excited to level 2 transit to lower level by emitting radiation $\nu_{ul}$, since collisional de-excitation is ineffective ... the condition for the formation of forbidden lines is that the electron density is smaller than the critical density. \item $N_e \gg N^c_e$ (dense gas). Since $A_{ul} \ll C_{ul}N_e$, we have $Q_{ul} \approx N_l A_{ul} C_{lu}/C_{ul}$... we get $Q_{ul} = N_l A_{ul}(g_u/g_l)exp(-\chi_{ul}/kT)$. The atoms excited to level 2 fall to the lower level mostly through collisional de-excitation. Therefore, although the emissivity is proportional to $A_{ul}$, the radiation in not practically observable due to extremely low values of $A_{ul}$. \end{enumerate} }

{\bf What the authors fail to note is that $N_l A_{ul} (g_u/g_l)exp(-\chi_{ul}/kT)$ is higher than $N_l N_e C_{ul}$. As a result the text leaves the reader with the impression that to emit a forbidden line a low density is required.} 

\item \cite{1980pim..book.....D}:
{\it However, there is a competing process which can suppress the radiation. The excited level can be de-excited by to a lower level by electron collision. ... Under laboratory conditions, this latter process dominates and photon emission is drastically reduced. On the other hand, the gas in the nebulae is so tenuous that radiative decay generally dominates and photons are readily emitted.}

\item \cite{Allen1987}:
{\it Electron impacts can also remove the energy before a photon is emitted, just as in the terrestrial laboratory. At high density, therefore, the forbidden lines are quenched.}

{\bf These are reflecting the Bowen interpretation with which we disagree.}

\item \cite{2018MNRAS.479.4153Y}:
{\it The absence of forbidden broad lines suggests that the BLR consists of dense regions that suppress the formation of these lines through collisional de-excitation of the gas.}

{\bf Collisional de-excitation does not suppress the formation of these lines as it is readily compensated by collisional excitation.}

\item \cite{1988PASP..100..412O} would appear to put a nail in the coffin of quenching with his figure 3 (reproduced here with the original caption). 

{\it As an example, imagine a hypothetical forbidden line at $\lambda 5000$, with transition probability $A_{ul} = 1.8\times10^{-2} s^{-1}$ and $C_{ul} = 1.8\times10^{-8}\,cm^3\,s^{-1}$ at $T= 10^4$ K. These values approximately match [O{\small III}] $\lambda 5007$ and are typical of many forbidden lines observed in nebulae. Figure 3 shows the calculated emission rate (also taking $g_u/g_l = 1$ for simplicity). Now consider a hypothetical permitted line with all parameters the same, except the transition probability $A_{ul} = 10^8 s^{-1}$, typical of lines such as Na{\small I} $\lambda\lambda$5890,5896; Ca{\small II} $\lambda\lambda$3933,3968, etc. The emission rate would not change at low densities but would continue to increase with density to much higher values, as shown in Figure 3 also. The critical electron density for a line: $N_c = A_{ul}/C_{ul}$ at which the photon emission rate per ion goes over from increasing proportionally to electron density and becomes essentially constant, is $N_c \approx 10^6\,cm^{-3}$ for [O{\small III}] $\lambda\lambda$4959,5007 and other characteristic nebular lines, but is $N_c \approx 10^{15}\,cm^{-3}$ for strong permitted lines. Thus, in typical laboratory situations, with $N_e \ll 10^{15}\,cm^{-3}$, the forbidden lines are strongly collisionally de-excited and are weaker by factors  ~$10^9$ than the permitted lines and, hence, unobservable. On the other hand, at nebular densities they suffer little collisional de-excitation, and the long path lengths ($\approx0.1$\,pc in a planetary nebula) make them observable.}

\begin{figure*}[!htbp]
\centering
\includegraphics[width=4in]{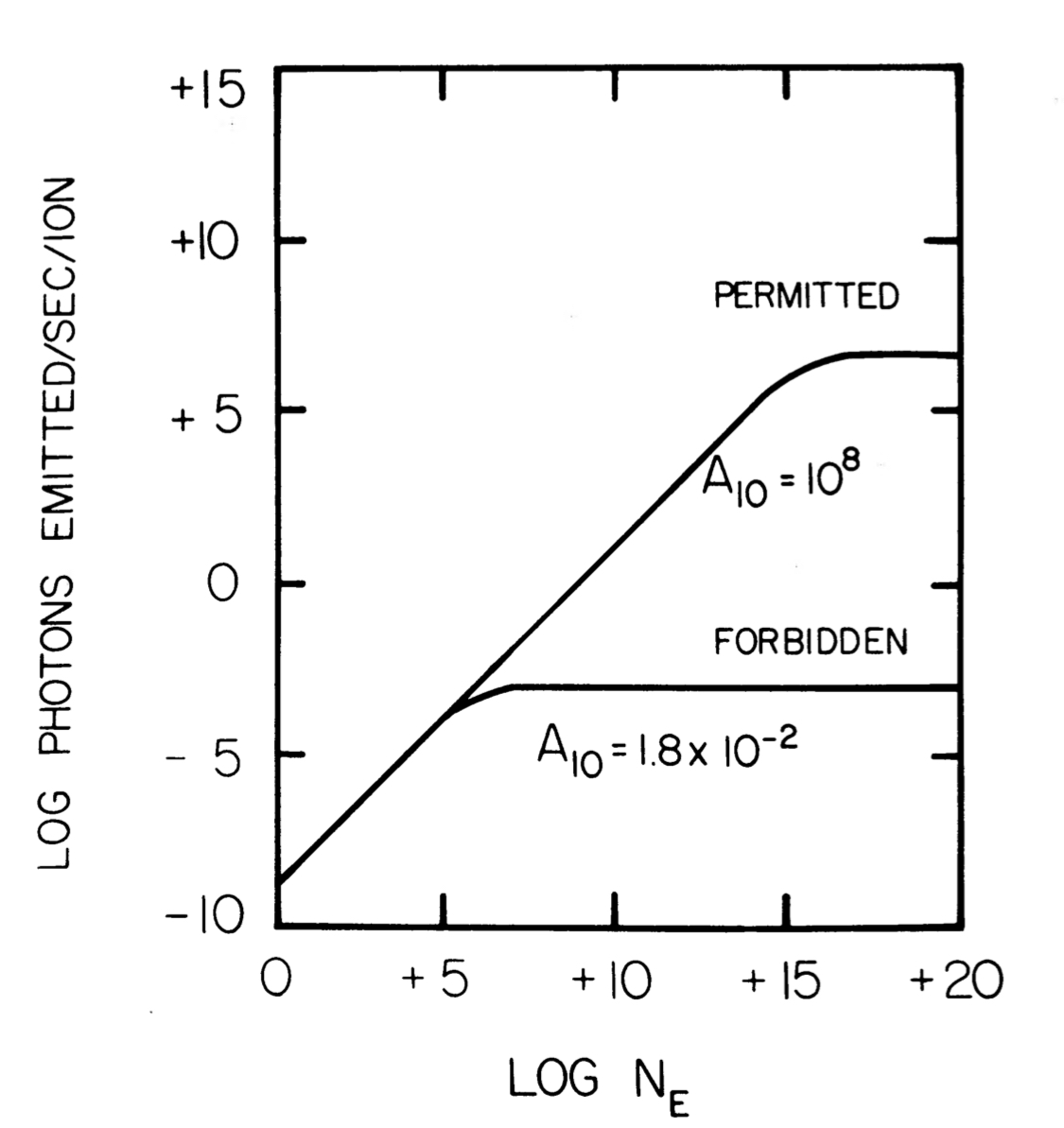}
\caption{Photo-emission rate per ion as a function of electron density for a typical forbidden line (lower curve) and a typical permitted line (upper curve) as explained in the text (\cite{1988PASP..100..412O} figure 3)}
\end{figure*}

{\bf The emissivity of the forbidden line is shown not to decline with increasing electron density. However, even here we cannot escape the words ``strongly collisionally de-excited'' when discussing unobservability.}

\item \cite{1995aelm.conf..134D}:
{\it [The infrared forbidden lines] are good density indicators for low-density regions, but ``saturate'' to asymptotic values at relatively low densities, whereupon they become good mass tracers.}  

{\bf Excellent! Crystal clear. This in agreement with our understanding. In the high density limit, the lines saturate.} 

\end{itemize}

\end{document}